\documentclass[review]{elsarticle}
\makeatletter
\def\ps@pprintTitle{%
 \let\@oddhead\@empty
 \let\@evenhead\@empty
 \def\@oddfoot{\centerline{\thepage}}%
 \let\@evenfoot\@oddfoot}
\makeatother
\usepackage{lineno,hyperref}
\modulolinenumbers[5]

\begin{document}

\begin{frontmatter}

\title{Effects of unsteady heat transfer on behaviour of Commercial Hydro-Pneumatic Accumulators}

\author[mymainaddress]{Jakob Hartig\corref{mycorrespondingauthor}}
\cortext[mycorrespondingauthor]{Corresponding author}
\ead{jakob.hartig@fst.tu-darmstadt.de}

\author[mymainaddress]{Benedict Depp}
\author[mymainaddress]{Manuel Rexer}
\author[mymainaddress]{Peter F. Pelz}


\ead[url]{www.fst.tu-darmstadt.de}
\address[mymainaddress]{TU Darmstadt, Chair of Fluid Systems, Otto-Berndt-Straße 2, 64287 Darmstadt}

\begin{abstract}
Hydraulic accumulators play a central role as energy storage in nearly all fluid power systems. The accumulators serve as pulsation dampers or energy storage devices in hydro-pneumatic suspensions. The energy carrying gas is compressed and decompressed, often periodically. Heat transfer to the outside significantly determines the transfer behaviour of the accumulator since heat transfer changes the thermodynamic state of the enclosed gas. The accumulators operating mode ranges from isothermal to adiabatic.
Simulating fluid power systems adequately requires knowledge of the transfer behaviour of the accumulators and therefore of the heat transfer.
The Engineer's approach to model heat transfer in technical system is Newton's law. However, research shows, that in harmonically oscillating gas volumes, heat flux and bulk temperature difference change their phase. Newton's law is incapable of representing this physical phenomenon.
We performed measurements on two sizes of commercial membrane accumulators. Experimental data confirm the failure of Newton's approach. Instead the heat transfer can be modelled with an additional rate dependent term and independently of the accumulator's size. Correlation equations for the heat transfer and the correct accumulator transfer behaviour are given.    

\end{abstract}


\end{frontmatter}


\section{Introduction}
\begin{figure}										
	\includegraphics[page =1, width=1\textwidth ]{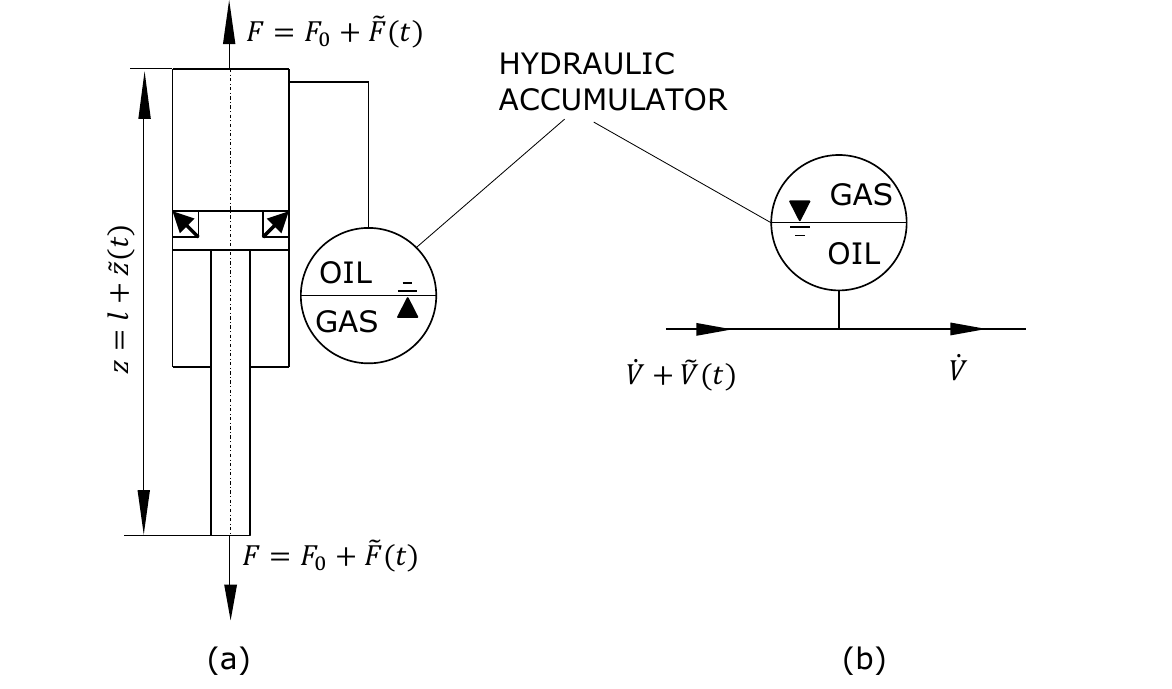}
	\caption{Hydraulic accumulators in a a) Hydropneumatic suspension strut, b) pulsation damper. In both cases the accumulator is excited with a time dependent volume flow rate.}			
	\label{fig:applications}					
\end{figure}

Hydro-pneumatic pressure accumulators - in short hydraulic accumulators - are used in oil-hydraulically operated machines and systems. Besides applications like energy storage or temperature compensation, hydraulic accumulators are used as potential energy storage in oscillating systems e.g. in hydro-pneumatic suspensions \cite{Bauer.2011} and for pulsation damping \cite{Zuti.2019} c.f. Figure \ref{fig:applications}. Carrier of potential energy is the periodically compressed working gas, usually nitrogen, which is undergoing a state change during compression and expansion \cite{Korkmaz.1982}. In analogy to mechanical oscillating systems, hydraulic accumulators act as springs. 
The state-change ranges from being isothermal to being adiabatic, where the nature of the state-change is mainly dependent on excitation frequency and the accumulator's size due to heat transport phenomena. Both dependencies can be captured by the ratio of two time scales: i) the duration of the compression cycle $2 \pi/\Omega$ and ii) the characteristic time of heat conduction in the accumulator $t_\mathrm{heat}$. The ratio of both time scales is a Péclet-number $Pe_\Omega$ which can be seen as a dimensionless frequency.
\begin{equation}
    Pe_\Omega =t_\mathrm{heat}\Omega /2\pi
\end{equation}
For $Pe_\Omega \to 0$ the state-change is isothermal ($n =1$), whereas for $Pe_\Omega \to \infty$ the state change is adiabatic ($n =1.4$) \cite{Pelz.2004}. 

Hydro-pneumatic suspensions and vibration absorbers are components of larger technical systems. For reliable system function and adequate controller design or harshness calculations, these systems have to be modelled adequately but concisely \cite{Isermann.2008}. There is ongoing research on adequate dynamic modelling of hydraulic accumulators in vibration applications regarding state equations \cite{vanderWesthuizen.2015}, operating temperature \cite{Els.1999} and heat transfer \cite{Pourmovahed.1984, Pourmovahed.1990}.


\begin{figure}																						
	\includegraphics[ width=1\textwidth ]{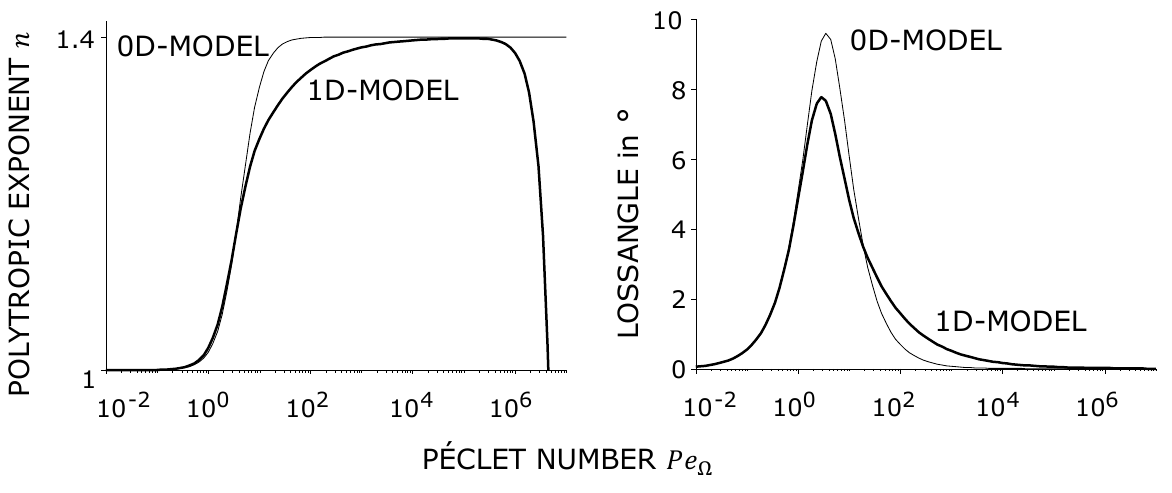}  
	\caption{Polytropic exponent and lossangle for oscillating gas volume under periodic excitation, with (1D-Model) and without (0D-Model) consideration of boundary layer formation. The models are from \cite{Pelz.2004}.}		
	\label{fig:comparison_polytropic}					
\end{figure}

Lumped parameter models are commonly used to model hydraulic accumulators. In lumped parameter models spatial dependency of thermodynamic quantities in the gas is neglected (0D-model). Consequently heat transfer is often assumed to be proportional to temperature difference between averaged bulk gas temperature $\bar{T} = 1/V \int T\mathrm{d}V$ and ambient temperature $T_a$ and therefore governed by Newton's law e.g.~\cite{Pourmovahed.1984,Rotthauser.15.01.1993,vanderWesthuizen.2015, Ho.2010}.
However, results from different fields show (see Sec.~\ref{sec:1dmodel} and Fig.~\ref{fig:comparison_polytropic}) that, the application of Newton's law in a lumped parameter model is questionable since the heat transfer in harmonically excited gas volumes is strongly influenced by the formation of boundary layers~\cite{Kornhauser.1994}. The polytropic exponent in Fig.~\ref{fig:comparison_polytropic} for the 1D-model with boundary layers is significantly lower in some frequency ranges compared to models in the literature \cite{Pourmovahed.1984,Rotthauser.15.01.1993,vanderWesthuizen.2015, Ho.2010}. Thus, the assumption of adiabatic state change is unacceptable for a larger frequency range than the models with Newton's law let assume.

This is especially relevant for system design, since the transfer behaviour, i.e. the change in pressure $p$ in relation to change in volume $V$ of the accumulator highly depends on the polytropic exponent $n$
\begin{equation}
    \frac{\mathrm{d}p}{\mathrm{d}V} = n \frac{p_1^2}{p_0 V_0},
    \label{eq:stiffness}
\end{equation}
where $p_1$ is the loading pressure, $p_0$ is the pre-charge pressure and $V_0$ is the initial gas volume \cite{Pelz.2004, Rexer.2020}. 
Therefore, in this paper we are dealing with the following question:

\emph{How can the behavior of hydraulic accumulators be modelled adequately but concisely?}

To answer this question a literature overview on heat transfer in periodically compressed gas volumes is given first. Then measurements of heat transfer in diaphragm hydraulic accumulators are presented and discussed. Furthermore the measurements will be used for tuning a semi-empiric lumped parameter model for heat transfer. Finally a comparison of the model and the standard model for hydraulic accumulators, found in the literature, is made.

\section{Review of Heat transfer in periodically compressed gas volumes}
\label{sec:1dmodel}

As seen with Eq.~\ref{eq:stiffness}, the transfer behaviour of hydraulic accumulators is highly dependent on the nature of the gas' state change and therefore on the heat transfer between gas and the environment ($\dot{Q}$). In the literature on hydraulic accumulators lumped parameter models with either adiabatic state change ($\dot{Q} = 0$) or Newtonian heat transfer ($\dot{Q} \propto \Delta T$) are used (c.f. \cite{Bauer.2011} or \cite{Pourmovahed.1984,Rotthauser.15.01.1993,vanderWesthuizen.2015, Ho.2010} respectively). In the latter case the heat flux $\dot{Q}$ is assumed to be proportional to the temperature difference of the gas bulk temperature $\bar{T}$ to the environment $\Delta T = (\bar{T}-T_a)$. The proportionality is usually captured with a heat transfer coefficient $\alpha$, where $\alpha$ is dependent on the fluid flow. The heat $\dot{Q}$ transferred to the surrounding gas with temperature $T_a$ is assumed to yield

 \begin{equation}
 \label{eq:Newton_Heatflux}
 \dot{Q}= - \alpha (\bar{T}-T_a)A,
 \end{equation}   
where $A$ is the heat transfer surface, $\bar{T}$ is the gas temperature and $\alpha$ is the heat transfer coefficient. The sign is so that $\dot{Q} > 0$ when heat is flowing from a hotter environment to the system. Even for many stationary cases heat transfer coefficients cannot be calculated theoretically and therefore correlations are tabulated, i.e. \cite{kind2010vdi}. To account for a wider range of problems, this is done in the form of Nusselt numbers $Nu := \alpha L/\lambda$, where $L$ is a specific length of the problem an $\lambda$ is the fluid heat conductivity.

Although not considered in hydraulic systems, the dynamic case of heat transfer in harmonically compressed gas volumes,  is relevant in different research fields namely i) cavitation bubbles, ii) air springs, iii) reciprocating piston engines. In the latter the heat from compression cannot be entirely separated from the combustion heat. Nevertheless the models can aid in understanding heat transfer phenomena at work. A literature review on these three research areas is presented in the following.

Pfriem \cite{Pfriem.1940} was the first to show with a theoretical analysis of a one dimensional model for piston engines that the heat transfer in periodically compressed gas volumes is frequency dependent due to boundary layer formation. In addition to that he proved that heat flux and wall bulk temperature difference are out of phase. Pfriem formulated the energy balance for a differential element of the boundary layer near the wall and obtained a convection-diffusion equation for the transported heat, which was solved by a perturbation approach for small pressure fluctuations. Similar models have been solved to model diesel engines \cite{Elser.1955}, cavitation bubbles \cite{Plesset.1977,Pelz.2009} and air springs \cite{lee1983simplistic,Pelz.2004}.

\begin{figure}																						
	\includegraphics[width = 1\textwidth]{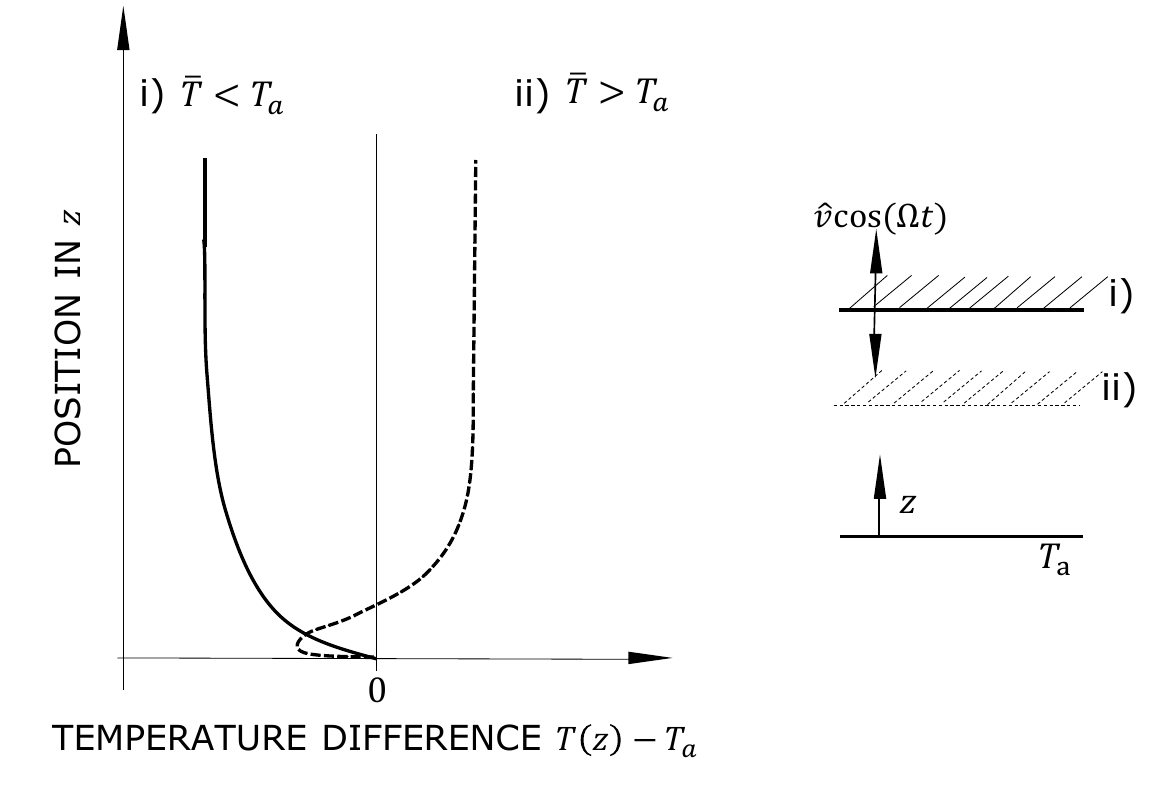} 
	\caption{Emergence of boundary layers during compression. Qualitative comparison of local temperature difference and bulk temperature difference at i) lower turning point and ii) a point during compression}			
	\label{fig:boundary_layers}					
\end{figure}

In his work Pfriem argues that fluid elements in the boundary layer near the wall dissipate the compression heat faster than fluid elements inside the gas. This idea is illustrated in Fig.~\ref{fig:boundary_layers} with a one-dimensional model, where the top moving wall is adiabatic and the lower one has the constant temperature $T_a$. First, at $t=0$ the gas is at its largest volume. The temperature profile of the temperature difference $\Delta T_{\mathrm{local}} = T(z) - T_a$ is as shown. The temperature $T(z)$ of fluid elements near the wall is strongly influenced by the wall temperature $T_a$. The local temperature gradient at the wall leads to a local heat flux $\dot{Q}_{\mathrm{local}}$ in positive $z$-direction, where heat is transferred from the warmer wall to the cooler fluid. Averaging $T(z)$ over the cross-section results in $\bar{T}$ that is lower than the constant wall temperature $T_a$. With the global temperature difference $\Delta T_{\mathrm{global}}= \bar{T} - T_a$ being negative, heat is predicted to flow from the warmer wall into the cooler fluid as well.
		
During compression the moving wall supplies energy to the gas in the form of volume work, which heats the gas over the entire cross-section. At point ii) of the cycle, $\bar{T}$ is already greater than $T_a$. However, the temperature of the gas near the wall does not change as quickly as in the bulk. $\nabla T$ in the zone near the wall still indicates heat flow from the locally warmer wall. However, the global temperature difference already predicts an opposite direction of heat flow. All in all, there is a phase difference in actual heat flux and global temperature difference. The Newtonian approach fails. As the compression progresses further, the temperature profile is monotonous again, so that the directions of global and local heat flux coincide.

Although the authors do not always state explicitly, all models mentioned above show that the heat transfer coefficient $\alpha$ during periodic compression only coincides with the heat transfer coefficient for stationary heat transfer in the case of small frequencies. For larger frequencies the models for periodic heat transfer predicts a phase difference between the heat flow and the driving temperature difference. It can be concluded that local heat flux and global temperature difference are transient and show a phase difference in periodically compressed gas volumes. So far only flows in one dimension were considered. In more than one dimension secondary flows occur and the heat fluxes vary over the surfaces in cylinders \cite{Willich.2017}.

Measurements confirm the theoretical results from above for internal combustion engines \cite{Annand.1980, Lawton.1987}. Measurements \cite{Kornhauser.1993, Kornhauser.1994} on gas springs and results from numerical models \cite{Lekic.2011} prove that heat flux and the bulk gas-wall temperature difference are out of phase.

The shortcomings of Eq. \ref{eq:Newton_Heatflux} for transient heat transfer can be overcome by extending the heat transfer term to be frequency dependent. In addition to that, the phase shift of heat flux and bulk temperature has to be addressed. Different authors considered the rate of change of the gas temperature to account for that during fitting of measurement results e.g. \cite{Annand.1970, Kornhauser.1994}. Another way to do that, is the complex extension of heat transfer $\alpha$, heat flux $\dot{Q}$ and temperatures $T$ \cite{Kornhauser.1989}. For harmonic oscillations the real part of the heat transfer then can be simplified to
\begin{equation}
\label{eq:Heatflux_New}
    \mathcal{R}(\underline{\dot{Q}}) = - \alpha'(\bar{T}-T_a)A - \frac{\alpha''}{\Omega} \frac{\mathrm{d}\bar{T}}{\mathrm{d}t}A.
\end{equation}

To date,all measurements and nearly all theoretical analyses were performed for cylindrical components. In cylindrical components, the compression is perpendicular to the main direction of heat transfer. All measurements were done for enclosures with one type of material (metal).

\section{Method}
\label{sec:method}
\begin{figure}																		
	\includegraphics[width = 1\textwidth]{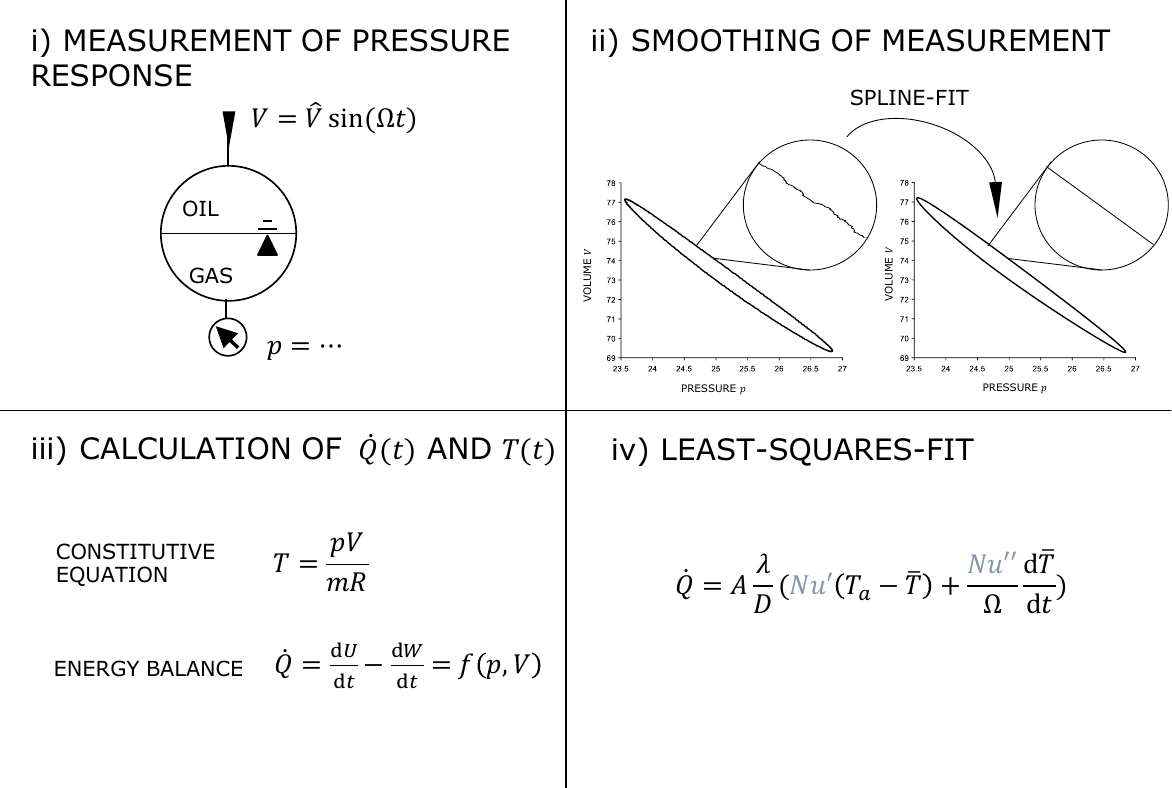} 
	\caption{Method for measuring and fitting heat flux and bulk temperature in commercial hydraulic accumulators}			
	\label{fig:method}					
\end{figure}

To answer the research question, measurements of the heat flux in commercial hydraulic accumulators have to be done. In this section an overview of the method for the investigation of heat flux, based on time varying pressure and volume measurements, is presented, c.f. Fig.~\ref{fig:method}. Using pressure and volume measurements, heat flux and bulk temperature is calculated with the help of energy equation and constitutive relations.

In Fig.~\ref{fig:method}, the first step i) are measurements of the pressure response of the hydraulic accumulators. For this a test rig was built so that the accumulators could be loaded with different pre-charge and load pressures and excited with a sinusoidal varying flow rate. The measurement data was then smoothed ii) since in step iii) derivatives are calculated. In step iii) constitutive relations for the gas and the momentary energy balance are used to calculate heat flux and bulk temperature. Finally, heat flux and temperature data are used to fit parameters in Eq. \ref{eq:Heatflux_New}.

\subsection{Test Rig and Test Objects}

\begin{figure}																		
	\includegraphics[width = 1\textwidth]{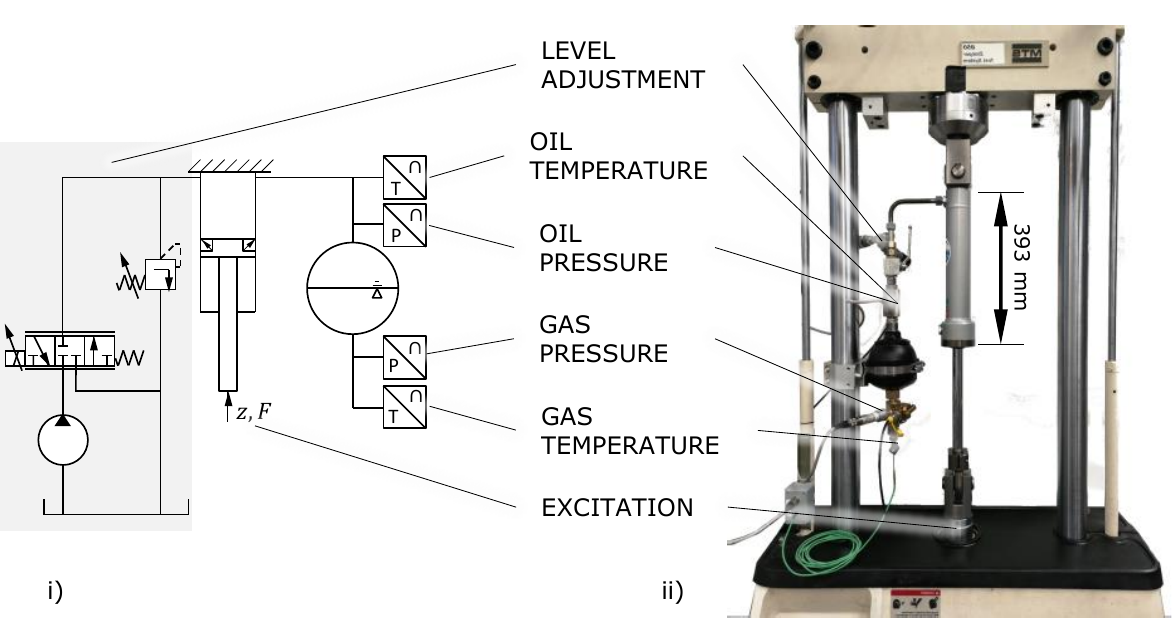} 
	\caption{Schematics and Test Rig}			
	\label{fig:test_rig}					
\end{figure}

The test-rig for commercial hydraulic accumulators was first presented in \cite{Rexer.2020}. Nevertheless some details on the test-rig are presented below. The basic design is similar to a hydro-pneumatic suspension strut (see Fig.~\ref{fig:applications}). A single-acting hydraulic cylinder is displaced in a path-controlled manner. Thus, oil volume is passed into the accumulator and the resulting pressure in the gas is measured (calibrated pressure transducer with measurement uncertainty of $0.0254 \, \mathrm{bar}$). In addition to that the system is level controlled, where the loading pressure can be adjusted by means of a 3-3-way valve. The real test structure and schematic diagram can be seen in Fig.~\ref{fig:test_rig}.  

The volume flow is based on the measured displacement of the cylinder $z_\mathrm{cyl}$ (uncertainty $0.5 \, \%$ of measurement value) and evaluated by multiplication with the area of the cylinder with $A_\mathrm{cyl}=1963.5\,\mathrm{mm}^2$. Volume flow in cylinder and accumulator is the same, since pipes and fittings only act as resistance and inductance between the cylinder and accumulator. The capacity of the piping is negligible up to 120 bar load pressure since the maximum stiffness of accumulators $k_\mathrm{acc} = \mathrm{d}p/\mathrm{d}V \approx 10^{10} \, \mathrm{N/m^5}$ is two orders of magnitudes lower than the stiffness for oil Shell Tellus S2 HLP 46 $k_\mathrm{oil} = 1.6 \cdot 10^{12} \, \mathrm{N/m^5}$ and hydraulic hoses $k_\mathrm{hose} = 5.66 \cdot 10^{12} \, \mathrm{N/m^5}$. \cite{Findeisen.2015}

The deviation of the excitation from a sinusoidal signal is negligible since the form factor 
\begin{equation}
 X_f=\frac{RMS}{AVR}   
\end{equation}
where RMS is the root mean square and AVR the mean value of the absolute value of the signal is sufficiently small. At the highest excitation (10~mm, 10~Hz) and the highest load pressure (120~bar) considered here, the form factor is $X_f=1.1149$ which corresponds to a deviation of $0.3793\,\%$ from the form factor of a sine signal. The form deviation is therefore negligible.

Both cylinder displacement and pressure signal are recorded in the same data acquisition device (dSpace DS1103) at 1 kHz. Phase lag between the two signals is therefore much lower than the measurement frequencies.

 Two sizes of diaphragm hydraulic accumulators  were measured (HYDAC SBO500-0,1A6/112U-500AK and SBO330-0,75E1/112U-33AB030) at frequencies from 0.002 Hz to 10 Hz, at different amplitudes and pre-charge and load pressures c.f. Table \ref{tab:test_plan}. The gas used was nitrogen and fulfilled the specifications of DIN EN ISO 14175: N1 (type 5.0 for 0.1 l accumulator with $99.999 \, \%$ purity and type 4.6 for 0.75 l accumulator with $99.996 \, \%$ purity)

\begin{table}
\begin{tabular}{ c c c }
        & HYDAC SBO500-0,1 & HYDAC SBO330-0,75 \\ 
nominal volume in l & 0.1  &  0.75 \\
heat-transfer surface $A$ in $10^{-3}$ m\textsuperscript{2} & 4.25& 20\\
 pre-charge pressures $p_0$ in bar & 20, 40 & 20, 40\\
 load pressures $p_1$ in bar & 25, 50 & 30, 60, 90, 120 \\
amplitude in ml & 4, 8 & 8, 20, 40 
\end{tabular}
\caption{\label{tab:test_plan} Test plan for hydraulic accumulators}
\end{table}
\begin{figure}																		
	\includegraphics[width = 1\textwidth]{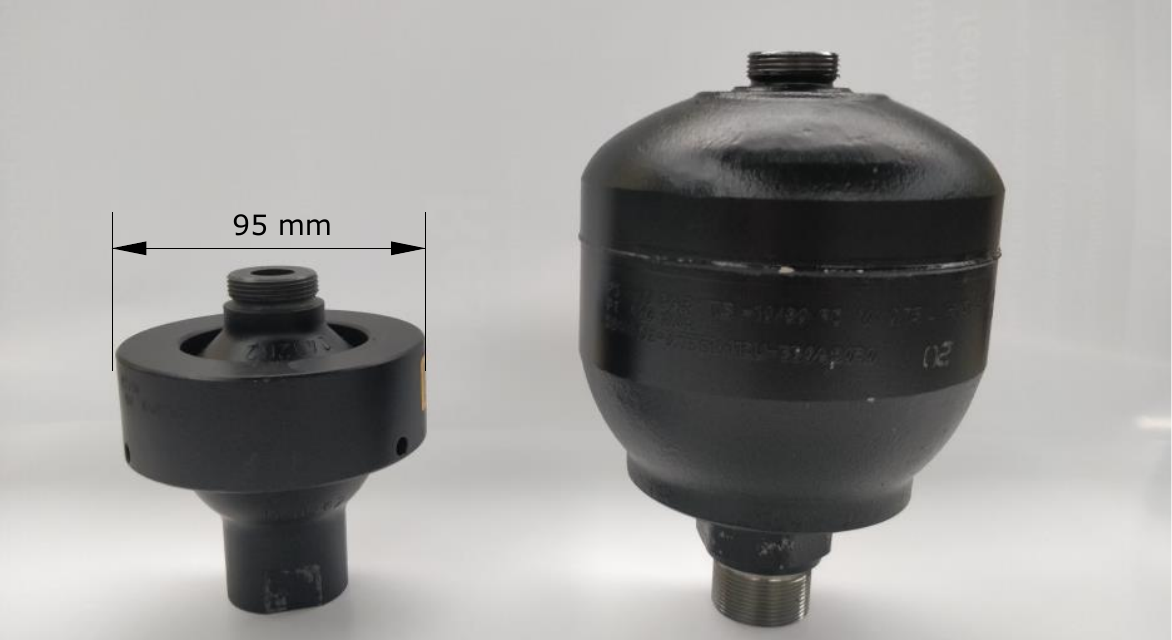} 
	\caption{Hydraulic accumulators used: HYDAC SBO500-0,1 (left) and SBOSBO330-0,75 (right)}			
	\label{fig:test_objects}					
\end{figure}

\subsection{Temperature and Heat Flux Calculation}
The pressure and flow rate measurements are used to calculate a momentary bulk temperature (the same temperature that is used in lumped parameter models), a momentary heat flux to the environment and the wall temperature of the accumulator. Unlike the temperature, there is no spatial pressure dependency expected, since the Mach-Number $Ma =0.003$ and standing waves would lead to pressure drop  \cite{spurk2007fluid}, \cite{Pelz.2004}. This method was first introduced by Kornhauser~\cite{Kornhauser.1989}.

The method has advantages in comparison to surface temperature and heat-flux measurements since the latter only show local heat-fluxes which may differ depending on location. In contrast to surface temperature measurements, the only assumption needed is that of constant pressure in the accumulator \cite{Kornhauser.1994}. The disadvantage of the method mentioned by Kornhauser, namely measurement noise influencing the temperature and heat flux calculations, was taken care of by cleaning the signals with the help of spline fitting.

To obtain the time history of the heat flow $\dot{Q}$ over the system boundary, the first law of thermodynamics for a closed system is applied to the hydraulic accumulator.
\begin{equation}
\delta Q = dE - \delta W
\label{eq:FirstLaw}
\end{equation}
The variation symbols indicate that $W$ and $Q$, unlike $E$, are inexact differentials.
The only form of work done is volume work, whose differential can be rewritten as $\delta W = -pdV$ with pressure $p$ and volume $V$.
The differential change of the internal energy of a mass $m$ of a calorically ideal gas is given by
\begin{equation}
\label{eq:internalenergy}
dE = m c_{v}dT.
\end{equation}
Also thermally ideal behavior is assumed. Throughout the experiments, nitrogen was used. Nitrogen behaves almost ideally at $300$ K to $100$ bar since the compressibility factor is very close to 1 \cite{Nelson.1955}. Therefore the ideal gas law in the form
\begin{equation}
pV = mRT
\label{eq:idGasdQ}
\end{equation}
can be used. Combining Eqs.~\ref{eq:FirstLaw} to \ref{eq:idGasdQ} results in an expression for $\delta Q $, which only depends on pressure and volume
\begin{equation}
\delta Q = \frac{\gamma}{\gamma - 1} p \, \mathrm{d}V + \frac{1}{\gamma - 1} V \, \mathrm{d}p.
\end{equation} 
Additionally the relations $\gamma= c_{p}/c_{v}$ and $R = c_{p} - c_{v}$ are used.  
Derivation and dividing by the heat transfer surface $A$ yields $\dot{q}$ for the heat flux density
\begin{equation}
\frac{1}{A} \frac{\delta Q}{\delta t} = \dot{q} = \frac{\gamma}{\gamma - 1}p\,\dot{V} + \frac{1}{\gamma - 1}V \, \dot{p}.
\label{eq:Kornhauserq}
\end{equation}
To calculate $\dot{q}$ from Eq. \ref{eq:Kornhauserq} the time signals of $p, \, V$ and their first time derivatives $\dot{p}, \, \dot{V}$ are necessary. Small fluctuations in the measurement data with a typical time scale much lower than the time scale of the oscillations would lead to large deviations in the derivatives. Therefore, the measurement data was spline fitted before fitting the Nusselt number.

Calculation of gas mass $m$ was done with the ideal gas law and temperature and pressure data from the stationary pre-charge state.

\subsection{Nusselt Fit}
The thermodynamic calculation of the previous section provides the time history of $\dot{q}$, $T(t)$ and $\dot{T}$. These quantities can be linked with each other by Eq.~\ref{eq:Heatflux_New}. 

However, more common is a dimensionless formulation of $\alpha$ with $Nu:=\alpha D_h/ \lambda$ in Eq.~\ref{eq:Heatflux_New}

\begin{equation}
\label{eq:fittingeq}
\dot{Q} =  A \frac{\lambda}{D_h} (Nu' (T_a -  \bar{T}) + \frac{Nu''}{\Omega}\frac{\mathrm{d}\bar{T}}{\mathrm{d}t}).
\end{equation}

The frequency $\Omega$ can be written dimensionless as a Péclet Number
\begin{equation}
Pe := \frac{\Omega D_h^2}{4 \lambda / (c_p \varrho_1)}.
\end{equation}

For one frequency the time series of $\dot{q}$-$T$-$\dot{T}$ defines the unknown real and imaginary parts of the Nusselt number $\mathit{Nu'}$, $\mathit{Nu''}$. Therefore a nonlinear least-square fit is performed with Eq.~\ref{eq:fittingeq} to determine the unknown parameters $Nu'$ and $Nu''$ averaged over one compression-expansion cycle.

\section{Results and Discussion}
The results from the measurements introduced in Sec.~\ref{sec:method} are discussed in this section. First the temperature and heat-flux calculations are discussed. After that the Nusselt-Fit is discussed.

\subsection{Temperature and Heat-Flux calculation}
Gas bulk temperature and heat flux were calculated from pressure and volume data. Thermodynamic consistency is given since the residuum of Eq.~\ref{eq:FirstLaw} is in the order of one per mille in relation to the calculated heat flux for all frequencies.

\begin{figure}
\centering
	\includegraphics{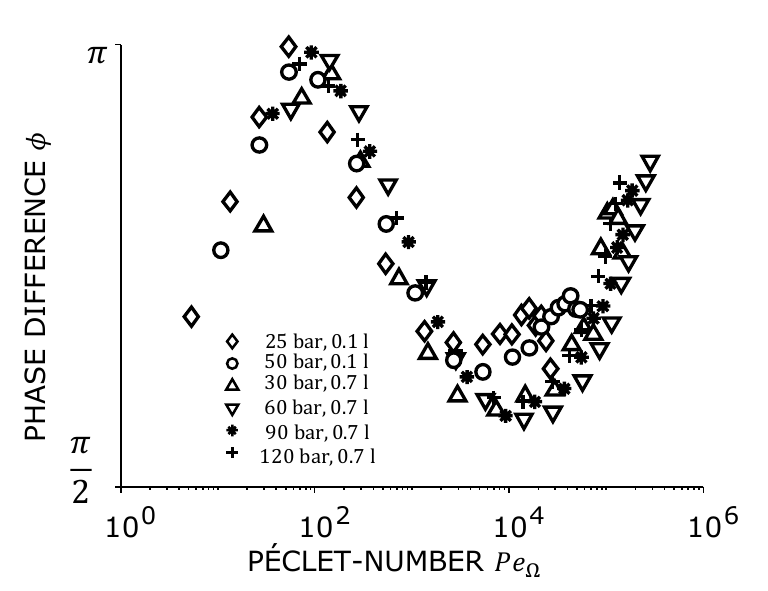} 
	\caption{Phase difference $\phi$ between mean bulk temperature $\bar{T}$ and heat flux $\dot{q}$ over Péclet-Number $Pe_\Omega$.}
	\label{fig:Phase_Results}		
\end{figure}

Theoretical considerations from Section~\ref{sec:1dmodel} predict a phase difference between the gas bulk temperature $\bar{T}$ and the heat flux $\dot{q}$. Calculating the phase difference between these two from the measurement data results in Fig.~\ref{fig:Phase_Results}. 
The phase difference of the different measurements correlates well with $Pe_\Omega$ despite different accumulator sizes and load pressures.

At low frequencies, in the isothermal region, when temperature rises so slowly that bulk temperature $\bar{T}$ is equal to the local temperature, heat-flux is negative and proportional to temperature. Therefore, expectations are, that heat-flux and temperature are out of phase at low frequencies with a phase difference of $\pi$. Then the phase difference decreases until it increases again.

In contrast to that, the measurements of $\phi$ have one maximum at $\pi$ and one minimum at $\pi/2$. Close inspection of the time course of volume and pressure measurements at the lowest measurement frequencies (0.002 Hz) for the 0.1~l accumulator shows a non-sinusodial behaviour where the $p$-$V$-hysteresis is not closed. The system was still in a transient state. Two causes may be the reason for that: i) Due to constraints in measurement time for the lowest frequencies only two instead of five full vibrations were measured. The two oscillations may be not enough to guarantee that the transient response has decayed. ii) The measurements were done in a non-climatised environment. Therefore fluctuations in environment temperature or radiant heat cannot be excluded at these long measurement times. In the following, the measurements at 0.002 Hz were excluded.

Nevertheless, a phase-difference is visible and failing of the Newtonian approach for heat transfer is evident for spatially averaged models of hydraulic accumulators.

\subsection{Nusselt-Numbers}

Nusselt Numbers were fitted from the bulk temperature and heat-flux calculations for two different accumulator sizes and different pre-charge pressures.
\begin{figure}									
	\includegraphics{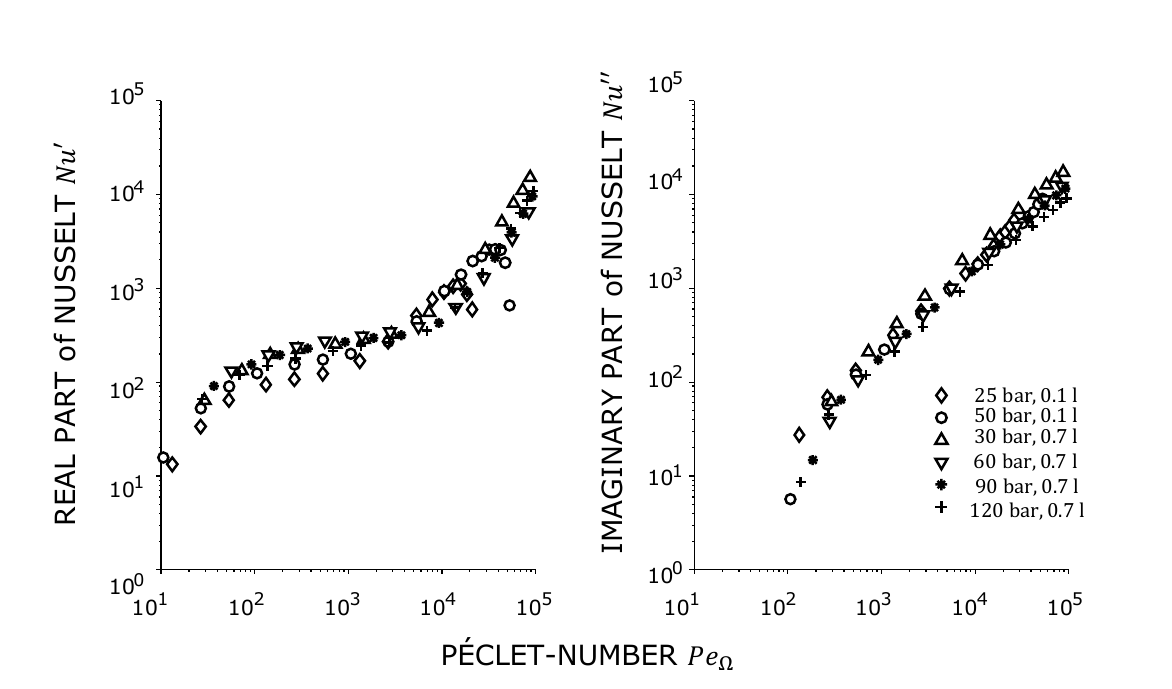} 
	\caption{Nusselt Fit for different Pressures and Accumulator sizes}	
	\label{fig:Nusselt_Fit}			
\end{figure}
The results in Fig.~\ref{fig:Nusselt_Fit} show good correlation with the dimensionless variables $Pe_\Omega$ and $Nu$ for the two different accumulators at different load pressures. The different volume and heat-transfer between the two accumulators seems to be captured by the hydraulic diameter $D$ and $Pe_\Omega$.

The trend of the real part of the Nusselt number $Nu'$ in Fig.~\ref{fig:Nusselt_Fit} can be divided into two parts with different slopes. The slope changes at about $Pe_\Omega=4\cdot10^3$.

For small $Pe_\Omega$ the real part $Nu'$ is larger than the imaginary part $Nu''$. Thus heat flow and temperature difference are exactly out of phase. For high $Pe_\Omega$, $Nu'\approx Nu''$ applies and the heat flow precedes the temperature difference by $\pi/2$.

\begin{figure}									
	\includegraphics{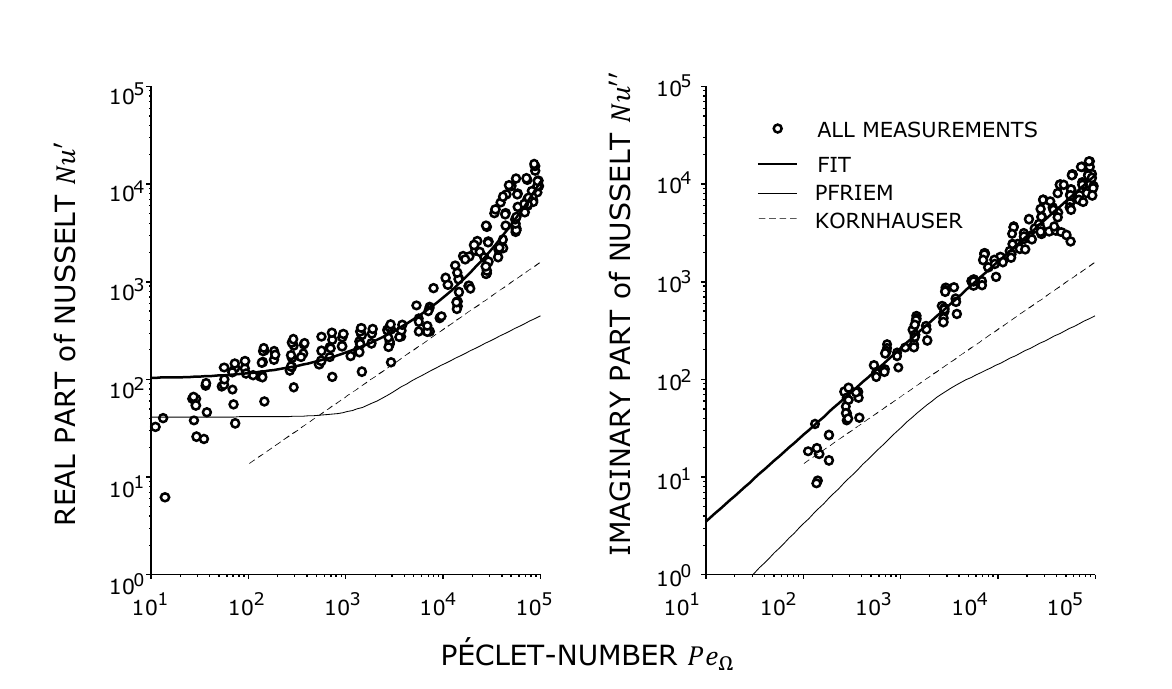} 
	\caption{All measured Nusselt-Numbers compared with fit from eq.~\ref{eq:fitnusseltreal} and \ref{eq:fitnusseltimag} and results from Pfriem \cite{Pfriem.1940} and Kornhauser \cite{Kornhauser.1994}}
	\label{fig:Nusselt_Fit_Korn}			
\end{figure}

 The division in two parts is visible in the model from Pfriem \cite{Pfriem.1940} and the measurements from Kornhauser \cite{Kornhauser.1994} (see Fig.~\ref{fig:Nusselt_Fit_Korn}). Compared to the results from Kornhauser~\cite{Kornhauser.1994}  however, the real part of Nusselt $Nu'$ is one order of magnitude larger and the Péclet-number, where the slope change occurs is about 1.5 magnitudes higher. We mainly attribute that to the non-similarity of the geometry of the two problems. The heat-transfer area and therefore the surface area to volume ratio has an enormous influence on the magnitude of $Nu$ and $Pe_\Omega$. 
The largest difference between the experiments described here and the measurements of Kornhauser is the rubber membrane in hydraulic accumulators compared to the full metal gas enclosure in Kornhauser's experiments. A Nusselt-fit with the full accumulators surface (metal + rubber) gave no correlation between the two different accumulator sizes. We therefore concluded, that the rubber surface must be excluded from the heat transfer. Comparing thermal diffusivity $a$ for metal ($5\cdot 10{-6} \, \mathrm{m^2/s}$) and NBR-rubber ($1.2\cdot 10{-7} \, \mathrm{m^2/s}$) leads to an order of magnitude difference between them. Consequently, the Fourier-number is one order of magnitude different for the two walls.

Another minor difference is the moving element in the gas enclosures: In Kornhauser's experiment a piston moved up and down, whereas in our experiments a rubber membrane moves in a complex way in more than one dimension. In contrast to parallel streamlines in pistons \cite{Lekic.2011} different primary and secondary flow patterns and therefore mixing of boundary layers at the rubber wall is probable to occur during movement.

The difference in heat transfer area due to the rubber-membrane and the general surface area to volume ratio difference, accounts for about 
\begin{equation}
Nu_\mathrm{acc}'\approx 5 Nu_\mathrm{Kornhauser}'
\end{equation}
and
\begin{equation}
Pe_{\Omega\mathrm{acc}}'\approx 7 Pe_{\Omega\mathrm{Kornhauser}}',
\end{equation}
compared with Kornhausers results. 



To use the data in models, a nonlinear least-square fit of the Nusselt-Numbers was done. The data-sets with smallest amplitude (4 ml for 0.1 l, 16 ml and 20 ml for 0.75 l) were used at different loading pressures. The data can be fitted with
\begin{equation}
\label{eq:fitnusseltreal}
 \log(Nu') =  2.011 +  0.003219\cdot \log(Pe_\Omega)^4,
\end{equation}
for the real part, where R-square = 0.9679  and root mean squared error = 0.1624 and 
\begin{equation}
\label{eq:fitnusseltimag}
  \log(Nu'') =  -0.3445 +  0.8862 \cdot \log(Pe_\Omega)
\end{equation}
for the imaginary part, where R-square = 0.9840  and root mean squared error = 0.1411. Thus the fitting equations account for most of the variation in the data (see for Fig.~\ref{fig:Nusselt_Fit_Korn}).



\section{Extended Lumped Parameter Model for Hydraulic Accumulators}

The frequency response of accumulators is highly dependent on heat transfer. In this chapter the lumped parameter model for hydraulic accumulators commonly found in the literature on fluid power systems is compared to an extended one. In the latter, the influence of periodic compression of gas on heat transfer is integrated. 

For the state change the two axioms mass conservation 
\begin{equation}
\label{eq:konti}
\varrho \frac{\mathrm{d} V}{\mathrm{d} t} + V \frac{\mathrm{d} \varrho}{\mathrm{d}t}   = 0
\end{equation}
and energy conservation 
\begin{equation}
	c_v V(\frac{\mathrm{d} T}{\mathrm{d}t} \varrho + T \frac{\mathrm{d} \varrho}{\mathrm{d}t})+ T \varrho c_p \frac{\mathrm{d} V}{\mathrm{d}t} = \dot{Q} ,
\end{equation}
need to hold where $V$ is the volume of the accumulator, $c_v$ and $c_p$ are the specific heat capacities of the gas. 

As stated above, we assume ideal behaviour for the gas 
\begin{equation}
\label{eq:ideal_gas}
p = \varrho R T,
\end{equation}
where $p$ is the pressure in the accumulator, $\varrho$ is the gas density, $R$ is the specific gas constant and $T$ is the temperature. However, all calculations in this paper can be done for non-ideal behaviour.

In our case the accumulator volume $V$ is changed dynamically, denoted by
\begin{equation}
\label{eq:excitation}
	V = V_0  +  \hat{V} \sin(\Omega t),
\end{equation}
where the index 0 denotes the pre-charged average working state of the accumulator. 
In the literature eq.~\ref{eq:Newton_Heatflux} is used for the heat flux $\dot{Q}$.

Linearization and transformation of equations \ref{eq:Newton_Heatflux} and \ref{eq:konti} to \ref{eq:excitation} into frequency space yields the $p/V$-transfer behaviour \cite{Pelz.2004}
\begin{equation}
\label{eq:transfer behaviour}
k^+ = \frac{\hat{p}^+}{\hat{V}^+} = - \frac{i \gamma Nu/Pe_\Omega - \gamma}{-i \gamma Nu/Pe_\Omega + 1},
\end{equation}
where $\hat{p}^+ = \hat{p}/p_0$ and $\hat{V}^+ = \hat{V}/V_0$. The hat denotes small deviations around the initial, pre-charged state. As used above, $Nu$ is the Nusselt number and $Pe_\Omega$ the dimensionless frequency.

Equation \ref{eq:transfer behaviour} is the transfer behaviour of hydraulic accumulators commonly found in the literature on fluid power systems (compare Fig.~\ref{fig:comparison_polytropic}).

Using eq. \ref{eq:Heatflux_New} instead of \ref{eq:Newton_Heatflux} for the heat-flux yields
\begin{equation}
    k^+ = \frac{\hat{p}^+}{\hat{V}^+} = - \gamma \frac{1 + 1/Pe_\Omega (Nu''-i Nu')}{\gamma/Pe_\Omega(Nu''-i Nu') + 1}
    \label{eq:transfer behaviour new}
\end{equation}
Equation \ref{eq:transfer behaviour new} is a generalization of equation \ref{eq:transfer behaviour}. In general $Nu$ is a function of $Pe_\Omega$. For this Eq. \ref{eq:fitnusseltreal} and \ref{eq:fitnusseltimag} were used.

\begin{figure}									
	\includegraphics{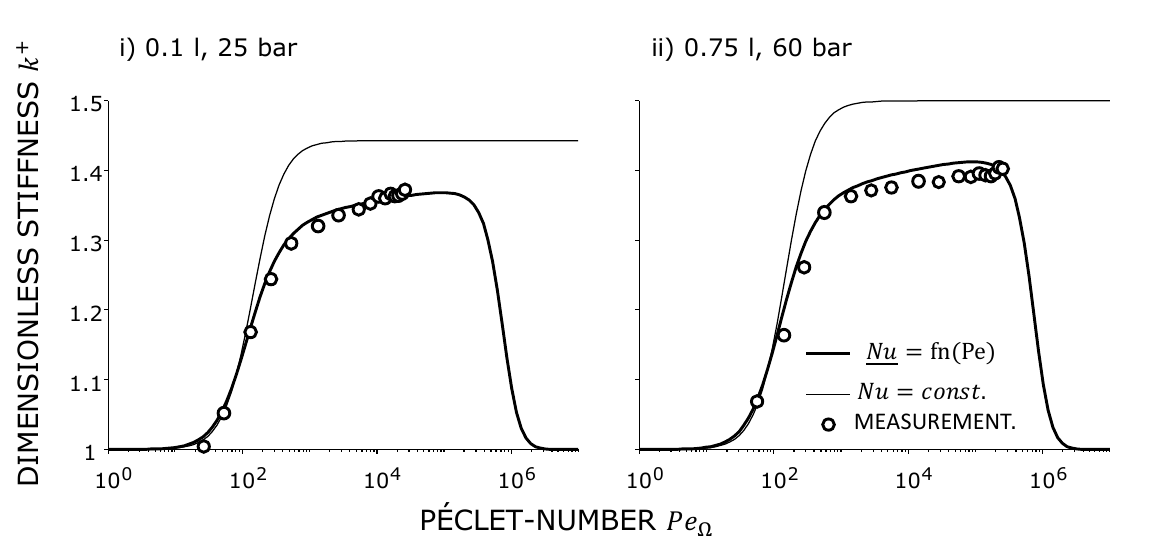} 
	\caption{Comparison of dimensionless stiffness $k^+$ of measurement, old model eq.~\ref{eq:transfer behaviour}  with $Nu=const.$ and new model eq.~\ref{eq:transfer behaviour new} with $Nu = \mathrm{fn}(Pe_\Omega)$ for i) 0.1 l accumulator and ii) 0.75 l accumulator}	
	\label{fig:stiffness}			
\end{figure}

In Fig.~\ref{fig:stiffness} the two models are compared with measurement data. The new extended model is significantly better than standard model.

Furthermore, the model seems to be useful for extrapolation, when compared with Fig.~\ref{fig:comparison_polytropic}. The drop in stiffness due to standing waves, as described in \cite{Pelz.2004}, is visible.

\section{Conclusion and Outlook}
It can be concluded, that the usual assumptions of accumulators being adiabatic or Newton's law for heat transfer are wrong for hydro-pneumatic accumulators for $Pe_\Omega \ge 10^2$.

In the measurement data for harmonically excited hydraulic accumulators a phase difference between heat flux and temperature is measurable.

For Pe-Number $Pe_\Omega \ge 10^2$ heat transfer in hydraulic accumulators can be modelled semi-empirically but universally for two different sizes of accumulators


\bibliography{adsorption_acc_bib}

\end{document}